\documentclass[prl,twocolumn,showpacs,floatfix]{revtex4}
\usepackage{amsmath}
\usepackage{amsfonts}
\usepackage{graphics}

\newcommand{\vect}[1]{\boldsymbol{#1}}

\newcommand{\figwidth}{7.75cm}

\begin{document}

\title{Fractional orbital occupation of spin and charge in artificial
  atoms}

\author{Jordan Kyriakidis}
\homepage{http://soliton.phys.dal.ca}
\author{Catherine J.~Stevenson}
\affiliation{Department of Physics and Atmospheric Science, Dalhousie
  University, Halifax, Nova Scotia, Canada, B3H 3J5}

\begin{abstract}
  We present results on spin and charge correlations in
  two-dimensional quantum dots as a function of increasing Coulomb
  strength (dielectric constant).  We look specifically at the orbital
  occupation of both spin and charge.  We find that charge and spin
  evolve separately, especially at low Coulomb strength.  For the
  charge, we find that a hole develops in the core orbitals at strong
  Coulomb repulsion, invalidating the common segregation of confined
  electrons into an inert core and active valence electrons.  For
  excitations, we find a total spin-projection $S_z = -1/2$ breaks
  apart into separate occupations of positive and negative spin.  This
  dissociation is caused by spin correlations alone.  Quantum
  fluctuations arising from long-range Coulomb repulsion destroy the
  spin dissociation and eventually results in all orbitals carrying a
  negative spin.
\end{abstract}

\date{\today} 

% PACS
% Quantum dots
%   devices, 85.35.Be
%   electron states of, 73.21.La
%   magnetic properties of, 75.75.+a
% Colective excitations
%   nanoscale systems73.21.-b
\pacs{85.35.Be, 73.21.La, 75.75.+a, 73.21.--b}

\maketitle

Quantum dots, especially at zero magnetic field, can often be
considered artificial atoms~\cite{mceuen97:artif.atoms,kastner00:SET}.
Like real atoms, artificial atoms can confine a definite number of
electrons, and the electronic and magnetic properties are largely
governed by the actual number of confined particles.  On the other
hand, there are dramatic differences between the two which stem
primarily and rather obviously from the synthetic nature of quantum
dots.  Quantum dots are not identical in the quantum-mechanical sense,
and they are more tunable~\cite{kouwen01:few.elect.quant.dots} than
most atomic and condensed matter systems.  Both the qualitative and
quantitative features of the confinement, for example, are tunable;
this, in turn, can be exploited to control the
angular-momentum~\cite{kyriakidis05:p_shell_hybrid} and the
spin~\cite{Kyriakidis02:Voltage-tunable-singlet-triplet,
  kogan03:singl.tripl.trans, Petta05:Coherent-Manipulation-of-Coupled}
content of ground and excited states.  With respect to the present
work, the most significant difference between atoms and
electrostatically-defined quantum dots in particular is their size.
Due to their much larger size, quantum dots typically have
single-particle level spacings of order 1~meV---approximately $10^4$
times smaller than atoms.  Consequently, the relative strength of
Coulomb repulsion is also $10^4$ times larger in quantum dots than it
is in atoms.  In these quantum dot systems, Coulomb effects are never
negligibly small and, for few electron systems especially, the
long-range nature of Coulomb repulsion often drives the observed
phenomena~\cite{ghosal06:_correl_induc_inhom}.

Our focus in this work is on the orbital occupation of particles.  We
seek to determine which orbitals are occupied by charge, which by
spin, and how much of each is contained in the given orbitals.  We
have investigated this question as a function of decreasing dielectric
constant, effectively increasing the strength of long-range Coulomb
repulsion.  We generally find that correlations can be induced both by
the spin symmetry as well as Coulomb effects.  Spin correlations are
most prominent at relatively weak interactions and we find in this
regime marked differences between the distribution of charge on the
one hand and spin on the other.  Coulomb fluctuations generally
destroy this spin-charge distinction, resulting in identical
distributions of both spin and charge.

We specifically report here on three main results.  First, the common
segregation of the charge into an inert core filling the low-lying
shells and an active valence shell completely breaks down at strong
interactions; the core shells are neither filled nor inert.  We find a
hole developing in the core as electrons maximise correlations by
occupying higher shells, depleting the core.  Second, we find that the
evolution of the fractional occupation of spin is different than that
of charge, making it useful to consider spin and charge as independent
entities even though they are both attached to each electron.
Thirdly, we look at excitations which preserve the symmetries of the
ground state and find significant correlation effects due entirely to
the spin symmetry, even in the complete absence of Coulomb effects.
We find that a spin excitation of $S_z = -\hbar/2$ does not merely
delocalize across the various orbitals, but rather dissociates into a
relatively large negative spin and a small positive spin, such that
the global spin symmetry is preserved.  Quantum fluctuations due to
long-range repulsion destroy this effect and lead to a single
delocalized spin with $S_z = -\hbar/2$.

Our model is a simple two-dimensional system of $N$ electrons confined
to a circular parabolic potential with long-range Coulomb repulsion.
The Hamiltonian is
\begin{multline}
  \label{eq:FockDarwin}
  \hat{H} = \frac{1}{2m} \left(
    \hat{\vect{p}} + \frac{e}{c}\vect{A} \right)^2 +
  \frac{1}{2} m \omega_0^2 \hat{\vect{r}}^2 \\ +
  \frac{\alpha}{2\varepsilon_{\text{GaAs}}} \sum_{i, j}^{(i \ne j)}
  \frac{e^2}{\left|\hat{\vect{r}}_i - \hat{\vect{r}}_j\right|},
\end{multline}
where the dielectric constant $\varepsilon = \varepsilon_{\text{GaAs}}
/ \alpha$ is scaled by $\alpha$ relative to that of GaAs
($\varepsilon_{\text{GaAs}} \approx 12.4$).  Thus, $\alpha = 0$
describes the non-interacting limit with Coulomb repulsion growing
linearly with $\alpha$, and with $\alpha = 1$ describing the GaAs
system.  Our results below are expressed as functions of $\alpha$.  In
all cases with $\alpha \ne 0$, we consider the full long-range ($\sim
1/r$) repulsion.

Our strategy is to first obtain the many-body eigenstates of
Eq.~(\ref{eq:FockDarwin}), including correlation effects.  We
subsequently analyze these many-body states (rather than just the
energies) and directly determine the orbital occupations of the
low-lying states.

The numerical approach we adopt is guided by three main requirements.
First, we require the full many-body states, properly accounting for
all exchange and correlation effects; techniques based on effective
single-particle orbitals or those based on obtaining energies only are
therefore insufficient.  Second, we also require excitations above the
the ground state.  Indeed we show below that an excitation of a single
$S_z = -1/2$ system results in both positive and negative spin
occupation of orbitals.  Thirdly, in these confined systems, the
strongest correlations often occur for small occupation numbers where
appeals to screening effects are invalid.

For these reasons the most appropriate numerical technique is a
configuration-interaction (exact diagonalization) approach.  This
affords the full many-body correlations of the ground state and the
excitations, and the few-body systems we consider are not beyond the
reach of techniques with a heavy computational load.

The Hamiltonian~(\ref{eq:FockDarwin}) is exactly solvable in the
noninteracting ($\alpha = 0$) limit~\cite{fock28, darwin30,
  jacak97:quant.dots}.  In this case, the single-particle eigenstates
are a pair of harmonic oscillators $|nm\rangle$, whose notation is
given in \cite{Kyriakidis02:Voltage-tunable-singlet-triplet}; the
quantum number $n$ is a Landau level index, whereas $m$ denotes a
guiding center.  The angular momentum of orbital $|nm\rangle$ is $L_z
= (n-m)\hbar$ and the energy at zero field of the same orbital is
$E_{mn} = \hbar \omega_0 (n + m + 1)$.

Including spin, then, our single-particle orbitals are denoted by
$|nms\rangle$.  We build the many-body basis states from
antisymmetrised tensor products of the single-particle eigenstates,
$|n_1m_1s_1, n_2m_2s_2, \ldots \rangle = c^\dagger_{m_1n_1s_1}
c^\dagger_{m_2n_2s_2} \ldots |\text{vac}\rangle$.

The various rotational symmetries of Eq.~(\ref{eq:FockDarwin}) imply
conservation of angular momentum projection $L_z$, spin projection
$S_z$, and spin magnitude $S^2$.  We use these quantum numbers to
label the correlated eigenstates of the interacting problem.  $L_z$
and $S_z$ do not induce correlations, but $S^2$ may induce
correlations.  Indeed we show this to be the case particularly for the
excitations even in the absence of Coulomb repulsion.  In this regime,
spin-correlations are strongest.

We denote by $|S_z, L_z, i\rangle$ the $i^{\text{th}}$ state
$|n_1m_1s_1, n_2m_2s_2, \ldots \rangle$ with fixed angular momentum
$L_z = \sum_k \hbar (n_k - m_k)$ and fixed spin projection $S_z =
\sum_k \hbar s_k / 2$ ($s_k = \pm 1$).  This state is a single
(uncorrelated) antisymmetrised state and is an eigenstate of
Eq.~(\ref{eq:FockDarwin}) in the absence of interactions.  Similarly,
by $|S, S_z, L_z, j\rangle$, we denote the $j^{\text{th}}$ spin
eigenstate with spin magnitude $\hbar^2S(S+1)$.  This state is in
general a correlated state
\begin{equation}
  \label{eq:SpinEig}
  \left| S, S_z, L_z, j \right\rangle = \sum_i \alpha_i^j \left|  S_z,
    L_z, i \right\rangle,
\end{equation}
whose correlations are due entirely to the spin symmetry.  This
correlated state is also an eigenstate of
Eq.~(\ref{eq:FockDarwin})---for $\alpha = 0$---and the spectra will
generally consist of degenerate subspaces.  Finally, the state $|E, S,
S_z, L_z, k\rangle$ denotes the $k^{\text{th}}$ eigenstate of
Eq.~(\ref{eq:FockDarwin}) with energy $E$, and with arbitrary
$\alpha$.  This state contains Coulomb correlations in addition to
those imposed by the spin symmetry.  It can be written as a coherent
superposition of the (correlated) spin eigenstates in
Eq.~(\ref{eq:SpinEig}):
\begin{equation}
  \label{eq:HEigs}
  \left| E, S, S_z, L_z, k \right\rangle = \sum_j
  \beta_j^k \left| S, S_z, L_z, j \right\rangle.
\end{equation}
These states will in general lift the degeneracies contained in the
spin-correlated states alone.

Our immediate objective is to find all $E$ and all $\beta_j^k$ in
Eq.~(\ref{eq:HEigs}).  In the configuration-interaction method,
Eq.~(\ref{eq:HEigs}) can be viewed either as a variational
\textit{ansatz} with the $\beta_j^k$ and the $E$ as variational
parameters, or, equivalently, as a direct matrix diagonalization in
the basis of the states $| S, S_z, L_z, j\rangle$, where the
$\beta_j^k$ and the $E$ emerge as the eigenvectors and eigenvalues of
the diagonalization procedure.  In either case, both views yield
identical results and both yield the exact solution in the limit of an
infinite number of terms in the sum of Eq.~(\ref{eq:HEigs}).

To construct our many-particle basis states, we start with a
sufficiently large set of single-particle orbitals $|nms\rangle$.  In
these calculations, we have taken 288 single-particle states.  For a
three-particle system with $S_z = -1/2$, this gives over $10^6$
possible three-body states.  From these, we consider only those
states, Eq.~(\ref{eq:SpinEig}), consistent with the desired quantum
numbers.  These symmetry requirements dramatically reduce the number
of basis states and, in practice, we find it rarely necessary to
consider greater than $\sim 10^3$ many-body states $|S_z, L_z,
i\rangle$.  From these states, we form all correlated spin eigenstates
$| S, S_z, L_z, j\rangle$, Eq.~(\ref{eq:SpinEig}), for a given quantum
number $S$.  The $\alpha_i^j$ can be determined by direct
diagonalization of the spin operator $\hat{S}^2$ in the $|S_z, L_z,
i\rangle$ basis~\cite{wensauer04:config.inter.method.fock}.
Alternatively, a closed, analytic, and exact form for the $\alpha_i^j$
can be determined given the number of singly- and doubly-occupied
orbitals using group theory methods and Clebsch-Gordon
technology~\cite{helgaker00:_molec_elect,rontani06:_full_config_inter}.

We use these computed correlated spin states, Eq.~(\ref{eq:SpinEig}),
as a basis to directly diagonalise the
Hamiltonian~(\ref{eq:FockDarwin}), thus obtaining the coefficients
$\beta_j^k$ in Eq.~(\ref{eq:HEigs}) as well as (less importantly in
the present case) the energies $E$.  A closed form expression for
arbitrary many-body matrix elements of the long-range Coulomb
interaction exists directly in spectral
space~\cite{Kyriakidis02:Voltage-tunable-singlet-triplet,
  rontani06:_full_config_inter}, obviating the need for real or
Fourier space integration.

For a given state $|\psi\rangle$ of the system, the charge occupation
(number of electrons) on the single-particle orbital $|nm\rangle$ is
given by
\begin{subequations}
  \label{eq:occ}
  \begin{equation}
    \label{eq:charge-occ}
    \rho_{\psi}^{nm} = \langle \psi | (c^\dagger_{nm\uparrow}
    c_{nm\uparrow} + c^\dagger_{nm\downarrow} c_{nm\downarrow}) |
    \psi\rangle,
  \end{equation}
  whereas the spin ($S_z$) occupation of the same orbital is
  \begin{equation}
    \label{eq:spin-occ}
    S_{\psi}^{nm} = \langle \psi | (c^\dagger_{nm\uparrow}
    c_{nm\uparrow} - c^\dagger_{nm\downarrow} c_{nm\downarrow}) |
    \psi\rangle.
  \end{equation}
\end{subequations}
Once the eigenstates have been computed, the expressions in
Eq.~(\ref{eq:occ}) are evaluated for each orbital $(m, n)$.

A three-particle system, at $B = 0$, is perhaps the simplest system
containing non-trivial spin and charge correlations.  Since an
odd-number of electrons has finite spin for all states, we can
determine how both spin and charge distribute themselves.  For weak
interactions, the third electron is the sole occupant of the second
shell; as Coulomb strength increases, charge correlations can be
expected to play an increasingly significant role in the ground state
and elementary excitations.  A three-particle system, even at fixed
orbital occupations~\cite{Kyriakidis05:Coherent-rotations,unpublished}
may form correlations entirely due to the spin-physics, independent of
Coulomb correlations.

At weak fields, the ground state is formed from the manifold of states
with definite total spin $S = 1/2$, spin projection $S_z = -1/2$, and
angular momentum $L_z = -1$ (all in units of $\hbar$).

\begin{figure}
\resizebox{\figwidth}{!}{\includegraphics{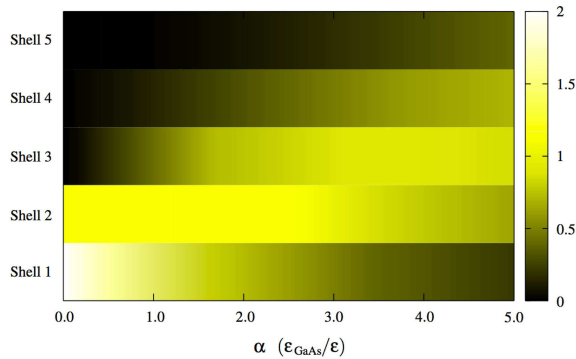}}
\caption{\label{fig:orbocc}Fractional orbital occupation of charge in
  the ground state.  The color-scale plot shows how the three
  particles distribute themselves across the various shells ($n+m$)
  at different interaction strengths.  With increasing interaction
  strength, a hole develops in the core orbital which, at zero
  interaction, is doubly occupied.  The plot depicts the
  three-particle ground state at zero field, $S=1/2$, $S_z = -1/2$,
  $L_z = -1$.}
\end{figure}

For the charge degrees of freedom, we have the general sum rule
$\sum_{mn}\rho^{mn}_{\psi} = N$, where $N=3$ is the number of confined
electrons.  At zero field, the single-particle spectrum consists of
degenerate shells, where the $k$th shell ($k = 1, 2, 3, \ldots$)
consists of all orbitals $|nm\rangle$ satisfying $n + m = k - 1$.
Figure~\ref{fig:orbocc} shows how the particles smear across the
shells for various interaction strengths.
At zero interactions, the exact ground state is a single
antisymmetrised state of a doubly occupied core (first shell) and a
single electron on the second shell~\cite{degeneracy}.  As a function
of increasing interaction strength, we find that the distribution
broadens.  This is expected since the orbital states are no longer
eigenstates of the Hamiltonian.  Surprisingly, however, the peak of
the distribution does not remain on the core shell, nor does the
occupation of the core drop to unity, thereby allowing exchange to
dominate.  Instead the core occupation drops below unity, almost to
zero, thereby leaving a hole in the core.  The electrons gain both
exchange and correlation energy by occupying higher shells at the
expense of single-particle energy.

\begin{figure}
  \resizebox{\columnwidth}{!}{\includegraphics{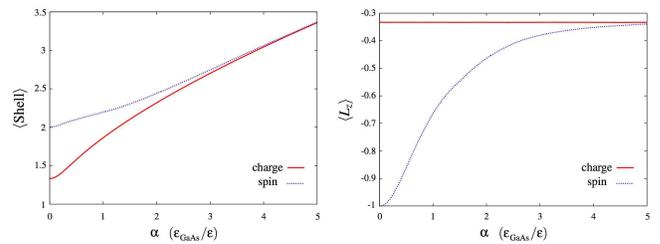}}
  \caption{\label{fig:spinEvolution}Separate spin and charge evolution
    with increasing interaction strength.  The panel at left displays
    the change in the mean shell occupied by spin and charge
    separately.  At right is shown the average angular momentum for
    both spin and charge.  At zero interactions, the spin distribution
    significantly differs from that of the charge, both among shells
    and angular momentum.  Coulomb-induced fluctuations destroy this
    order and restore congruence among charge and spin.}
\end{figure}

In addition to the charge occupation, we can look to the spin
occupation.  The ground state we are considering here has $S=1/2$ and
$S_z = -1/2$.  At zero interactions, the doubly occupied core shell
carries no spin; the spin sits in the second shell.  Similarly to the
charge, this excess spin delocalizes across the orbitals with
increasing interaction strength.  The evolution of the spin, however,
is distinct from that of the charge.  In Fig.~\ref{fig:spinEvolution},
the distinction is evident.  The left panel displays the mean shell
occupation for the ground state charge, $\sum_{m,n}(m+n)\rho^{mn}_0$,
and the ground state spin.  Also shown, in the right panel, is the
average angular momentum for both spin and charge.  For the charge,
the \emph{total} angular momentum is set by the global quantum numbers
to $L_z = -\hbar$ and so the average is fixed for all values of
$\alpha$.  These symmetry constraints do not apply to the (excess)
spin.  Instead, the spin obeys the sum rule $\sum_{m,n}s^{mn}_{\psi} =
-\hbar / 2$.

The distinction between spin and charge distributions is most evident
at weak interactions.  Here, only spin correlations play a prominent
role.  Upon increasing Coulomb strength, these additional fluctuations
destroy this distinction, resulting in identical mean shell and
angular-momentum occupations.

We turn now to excitations above the ground state.  In particular, we
look at excitations which preserve both angular momentum and spin.  We
find even stronger correlation effects due entirely to the spin
symmetry, independent of the Coulomb correlations.  To see this, we
look first at the noninteracting limit at zero field, and look at the
lowest excitations with the same quantum numbers as the ground state,
$(N, S, S_z, L_z) = (3, 1/2, -1/2, -1)$.

\begin{figure}
\resizebox{\figwidth}{!}{\includegraphics{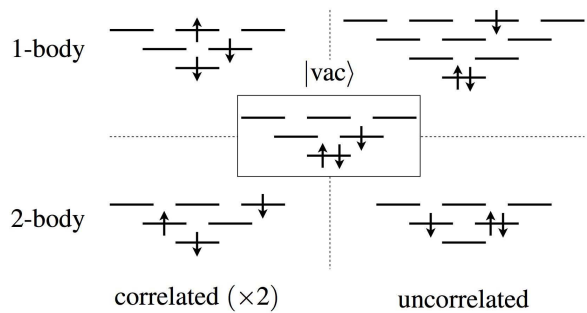}}
  \caption{\label{fig:excitations}Spin and angular momentum conserving
    excitations at zero interactions.  At centre is the ground state.
    The top row depicts one-body excitations and the bottom row
    two-body excitations.  The left column denotes spin correlated
    states (two per image) while the right denotes spin uncorrelated
states.}
\end{figure}

The lowest-energy excitations lie $2 \hbar \omega_0$ above the ground
state, and the subspace is six-fold degenerate.  These excitations can
be classified by two orthogonal criteria---whether they are a one-body
or a two-body excitation, and whether they are a spin correlated or
uncorrelated state.  The orbital configurations of these degenerate
excitations are depicted in Fig.~\ref{fig:excitations}.  The centre
image depicts the ground state (uncorrelated) configuration.  The top
set of plots shows the two one-body excitations, formed, in this case,
by moving a single particle up two shells.  (The shell spacing is
$\hbar \omega_0$.)  The bottom set depicts equally energetic
configurations, involving an excitations of two particles, each of
which moves up a single shell.  In contrast, the excitations can also
be classified according to their correlations.  The two images on the
right are uncorrelated states; after antisymmetrization, these states
are proper spin doublets ($S = 1/2$).  There are no additional degrees
of freedom due to spin.  the states on the left, however, are not
proper spin states even after antisymmetrization.  These
configurations, containing only singly-occupied orbitals, should be
viewed as each representing three states which differ solely by a
permutation of the spins.  From these three states, two orthogonal
$S=1/2$ states can be constructed.  (The third state has $S = 3/2$.)
These doublet states are correlated; they cannot be written as a
single antisymmetrised (Slater determinant) state in any basis.  The
correlations, however, are entirely induced by the spin-symmetry
requirements and not from Coulomb interactions.  (Indeed there are no
interactions at $\alpha=0$.)

\begin{figure}
\resizebox{\figwidth}{!}{\includegraphics{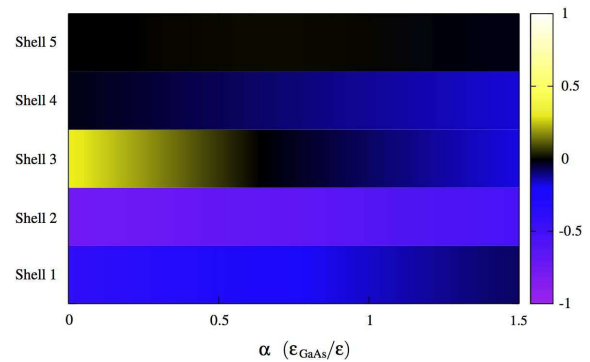}}
\caption{\label{fig:spinDissoc}Spin dissociation at zero interactions
  and its restoration with Coulomb repulsion.  In the presence of only
  spin correlations ($\alpha = 0$), the spin $S_z = -1/2$ breaks apart
  into a small positive component and a larger negative component,
  each occupying distinct regions of the quantum dot.  Coulomb
  correlations destroy this dissociation, resulting eventually in a
  single delocalized $S_z = -1/2$ particle.  The color bar is in units
  of $\hbar / 2$.}
\end{figure}

The implication of correlated excitations is that the excess spin ($Sz
= -1/2$) does not sit on a single orbital as in the ground state, even
when there is no interaction among the electrons.  In fact, it is not
even the case that the excess negative spin merely delocalizes.  This
single negative spin dissociates into a small positive spin and a
larger negative spin and these are each distributed on different
orbitals.  This dissociation can be traced back directly to the
correlated doublets in the degenerate manifold.  This is shown in
Fig.~\ref{fig:spinDissoc}, where we plot the net spin ($S_z$) per
shell as a function of increasing Coulomb strength.  A region of
positive spin exists in the third shell while all others have a net
negative spin.  (The \emph{total} spin is always $S_z = -\hbar / 2$.)
We stress that the dissociation of the spin is not due to Coulomb
correlations but is entirely due to the spin correlations.  In fact,
turning on the Coulomb correlations destroys the effect.  This is also
shown in Fig.~\ref{fig:spinDissoc}; the quantum fluctuations
associated with long-range Coulomb repulsion act to destroy the
positive spin contribution until all the remaining spin is only of one
species.

Correlation effects can be due to either long-range Coulomb repulsion
or spin symmetry.  At low fields, where low spin is often
energetically favoured, spin-correlation effects are strongest and
novel spin-induced phenomena may occur.  Coulomb fluctuations in
general induce different, often much stronger, effects and these
charge correlations often overwhelm, destroy, or otherwise mask
effects due to the spin symmetry.  We have shown here that in the
relatively simple and clean system of three interacting electrons with
both orbital and spin rotational symmetry the distinction and
dissolution of spin-symmetry-induced and Coulomb-induced phenomena can
be demonstrated at moderate Coulomb strength.

\begin{acknowledgments}
This work was supported by NSERC of Canada and by the Canadian
Foundation of Innovation.
\end{acknowledgments}

% \bibliography{bibabbrevs,orbocc,0D,misc}
%\bibliography{bibabbrevs,orbocc.custom}

\begin{thebibliography}{17}
\expandafter\ifx\csname natexlab\endcsname\relax\def\natexlab#1{#1}\fi
\expandafter\ifx\csname bibnamefont\endcsname\relax
  \def\bibnamefont#1{#1}\fi
\expandafter\ifx\csname bibfnamefont\endcsname\relax
  \def\bibfnamefont#1{#1}\fi
\expandafter\ifx\csname citenamefont\endcsname\relax
  \def\citenamefont#1{#1}\fi
\expandafter\ifx\csname url\endcsname\relax
  \def\url#1{\texttt{#1}}\fi
\expandafter\ifx\csname urlprefix\endcsname\relax\def\urlprefix{URL }\fi
\providecommand{\bibinfo}[2]{#2}
\providecommand{\eprint}[2][]{\url{#2}}

\bibitem[{\citenamefont{McEuen}(1997)}]{mceuen97:artif.atoms}
\bibinfo{author}{\bibfnamefont{P.~L.} \bibnamefont{McEuen}},
  \bibinfo{journal}{Science} \textbf{\bibinfo{volume}{278}},
  \bibinfo{pages}{1729} (\bibinfo{year}{1997}).

\bibitem[{\citenamefont{Kastner}(2000)}]{kastner00:SET}
\bibinfo{author}{\bibfnamefont{M.~A.} \bibnamefont{Kastner}},
  \bibinfo{journal}{Ann.\ Phys.\ (Leipzig)} \textbf{\bibinfo{volume}{9}},
  \bibinfo{pages}{885} (\bibinfo{year}{2000}).

\bibitem[{\citenamefont{Kouwenhoven et~al.}(2001)\citenamefont{Kouwenhoven,
  Austing, and Tarucha}}]{kouwen01:few.elect.quant.dots}
\bibinfo{author}{\bibfnamefont{L.~P.} \bibnamefont{Kouwenhoven}},
  \bibinfo{author}{\bibfnamefont{D.~G.} \bibnamefont{Austing}},
  \bibnamefont{and} \bibinfo{author}{\bibfnamefont{S.}~\bibnamefont{Tarucha}},
  \bibinfo{journal}{Rep.\ Prog.\ Phys.} \textbf{\bibinfo{volume}{64}},
  \bibinfo{pages}{701} (\bibinfo{year}{2001}).

\bibitem[{\citenamefont{Kyriakidis}(2005)}]{kyriakidis05:p_shell_hybrid}
\bibinfo{author}{\bibfnamefont{J.}~\bibnamefont{Kyriakidis}},
  \bibinfo{journal}{J.\ Phys.: Condens.\ Matter} \textbf{\bibinfo{volume}{17}},
  \bibinfo{pages}{2715} (\bibinfo{year}{2005}).

\bibitem[{\citenamefont{Kyriakidis et~al.}(2002)\citenamefont{Kyriakidis,
  Pioro-Ladriere, Ciorga, Sachrajda, and
  Hawrylak}}]{Kyriakidis02:Voltage-tunable-singlet-triplet}
\bibinfo{author}{\bibfnamefont{J.}~\bibnamefont{Kyriakidis}},
  \bibinfo{author}{\bibfnamefont{M.}~\bibnamefont{Pioro-Ladriere}},
  \bibinfo{author}{\bibfnamefont{M.}~\bibnamefont{Ciorga}},
  \bibinfo{author}{\bibfnamefont{A.~S.} \bibnamefont{Sachrajda}},
  \bibnamefont{and} \bibinfo{author}{\bibfnamefont{P.}~\bibnamefont{Hawrylak}},
  \bibinfo{journal}{Phys.\ Rev.\ B} \textbf{\bibinfo{volume}{66}},
  \bibinfo{pages}{035320} (\bibinfo{year}{2002}).

\bibitem[{\citenamefont{Kogan et~al.}(2003)\citenamefont{Kogan, Granger,
  Kastner, Goldhaber-Gordon, and Shtrikman}}]{kogan03:singl.tripl.trans}
\bibinfo{author}{\bibfnamefont{A.}~\bibnamefont{Kogan}},
  \bibinfo{author}{\bibfnamefont{G.}~\bibnamefont{Granger}},
  \bibinfo{author}{\bibfnamefont{M.~A.} \bibnamefont{Kastner}},
  \bibinfo{author}{\bibfnamefont{D.}~\bibnamefont{Goldhaber-Gordon}},
  \bibnamefont{and}
  \bibinfo{author}{\bibfnamefont{H.}~\bibnamefont{Shtrikman}},
  \bibinfo{journal}{Phys.\ Rev.\ B} \textbf{\bibinfo{volume}{67}},
  \bibinfo{pages}{113309} (\bibinfo{year}{2003}).

\bibitem[{\citenamefont{Petta et~al.}(2005)\citenamefont{Petta, Johnson,
  Taylor, Laird, Yacoby, Lukin, Marcus, Hanson, and
  Gossard}}]{Petta05:Coherent-Manipulation-of-Coupled}
\bibinfo{author}{\bibfnamefont{J.~R.} \bibnamefont{Petta}},
  \bibinfo{author}{\bibfnamefont{A.~C.} \bibnamefont{Johnson}},
  \bibinfo{author}{\bibfnamefont{J.~M.} \bibnamefont{Taylor}},
  \bibinfo{author}{\bibfnamefont{E.~A.} \bibnamefont{Laird}},
  \bibinfo{author}{\bibfnamefont{A.}~\bibnamefont{Yacoby}},
  \bibinfo{author}{\bibfnamefont{M.~D.} \bibnamefont{Lukin}},
  \bibinfo{author}{\bibfnamefont{C.~M.} \bibnamefont{Marcus}},
  \bibinfo{author}{\bibfnamefont{M.~P.} \bibnamefont{Hanson}},
  \bibnamefont{and} \bibinfo{author}{\bibfnamefont{A.~C.}
  \bibnamefont{Gossard}}, \bibinfo{journal}{Science}
  \textbf{\bibinfo{volume}{309}}, \bibinfo{pages}{2180} (\bibinfo{year}{2005}).

\bibitem[{\citenamefont{Ghosal et~al.}(2006)\citenamefont{Ghosal,
  G\"{u}\c{c}l\"{u}, Umrigar, Ullmo, and
  Baranger}}]{ghosal06:_correl_induc_inhom}
\bibinfo{author}{\bibfnamefont{A.}~\bibnamefont{Ghosal}},
  \bibinfo{author}{\bibfnamefont{A.~D.} \bibnamefont{G\"{u}\c{c}l\"{u}}},
  \bibinfo{author}{\bibfnamefont{C.~J.} \bibnamefont{Umrigar}},
  \bibinfo{author}{\bibfnamefont{D.}~\bibnamefont{Ullmo}}, \bibnamefont{and}
  \bibinfo{author}{\bibfnamefont{H.~U.} \bibnamefont{Baranger}},
  \bibinfo{journal}{Nature Phys.} \textbf{\bibinfo{volume}{2}},
  \bibinfo{pages}{336} (\bibinfo{year}{2006}).

\bibitem[{\citenamefont{Fock}(1928)}]{fock28}
\bibinfo{author}{\bibfnamefont{V.}~\bibnamefont{Fock}}, \bibinfo{journal}{Z.\
  Phys.} \textbf{\bibinfo{volume}{47}}, \bibinfo{pages}{446}
  (\bibinfo{year}{1928}).

\bibitem[{\citenamefont{Darwin}(1930)}]{darwin30}
\bibinfo{author}{\bibfnamefont{C.~G.} \bibnamefont{Darwin}},
  \bibinfo{journal}{Math.\ Proc.\ Cambridge Phil.\ Soc.}
  \textbf{\bibinfo{volume}{27}}, \bibinfo{pages}{86} (\bibinfo{year}{1930}).

\bibitem[{\citenamefont{Jacak et~al.}(1997)\citenamefont{Jacak, Hawrylak, and
  W{\'{o}}js}}]{jacak97:quant.dots}
\bibinfo{author}{\bibfnamefont{L.}~\bibnamefont{Jacak}},
  \bibinfo{author}{\bibfnamefont{P.}~\bibnamefont{Hawrylak}}, \bibnamefont{and}
  \bibinfo{author}{\bibfnamefont{A.}~\bibnamefont{W{\'{o}}js}},
  \emph{\bibinfo{title}{Quantum Dots}} (\bibinfo{publisher}{Springer},
  \bibinfo{address}{Berlin}, \bibinfo{year}{1997}).

\bibitem[{\citenamefont{Wensauer et~al.}(2004)\citenamefont{Wensauer,
  Korkusi\'nski, and Hawrylak}}]{wensauer04:config.inter.method.fock}
\bibinfo{author}{\bibfnamefont{A.}~\bibnamefont{Wensauer}},
  \bibinfo{author}{\bibfnamefont{M.}~\bibnamefont{Korkusi\'nski}},
  \bibnamefont{and} \bibinfo{author}{\bibfnamefont{P.}~\bibnamefont{Hawrylak}},
  \bibinfo{journal}{Solid State Commun.} \textbf{\bibinfo{volume}{130}},
  \bibinfo{pages}{115} (\bibinfo{year}{2004}).

\bibitem[{\citenamefont{Helgaker et~al.}(2000)\citenamefont{Helgaker,
  J{\o}rgensen, and Olsen}}]{helgaker00:_molec_elect}
\bibinfo{author}{\bibfnamefont{T.}~\bibnamefont{Helgaker}},
  \bibinfo{author}{\bibfnamefont{P.}~\bibnamefont{J{\o}rgensen}},
  \bibnamefont{and} \bibinfo{author}{\bibfnamefont{J.}~\bibnamefont{Olsen}},
  \emph{\bibinfo{title}{Molecular Electronic-Structure Theory}}
  (\bibinfo{publisher}{Wiley}, \bibinfo{address}{Chichester},
  \bibinfo{year}{2000}).

\bibitem[{\citenamefont{Rontani et~al.}(2006)\citenamefont{Rontani, Cavazzoni,
  Bellucci, and Goldoni}}]{rontani06:_full_config_inter}
\bibinfo{author}{\bibfnamefont{M.}~\bibnamefont{Rontani}},
  \bibinfo{author}{\bibfnamefont{C.}~\bibnamefont{Cavazzoni}},
  \bibinfo{author}{\bibfnamefont{D.}~\bibnamefont{Bellucci}}, \bibnamefont{and}
  \bibinfo{author}{\bibfnamefont{G.}~\bibnamefont{Goldoni}},
  \bibinfo{journal}{J.\ Chem.\ Phys.} \textbf{\bibinfo{volume}{124}},
  \bibinfo{pages}{124102} (\bibinfo{year}{2006}).

\bibitem[{\citenamefont{Kyriakidis and
  Penney}(2005)}]{Kyriakidis05:Coherent-rotations}
\bibinfo{author}{\bibfnamefont{J.}~\bibnamefont{Kyriakidis}} \bibnamefont{and}
  \bibinfo{author}{\bibfnamefont{S.~J.} \bibnamefont{Penney}},
  \bibinfo{journal}{Phys.\ Rev.\ B} \textbf{\bibinfo{volume}{71}},
  \bibinfo{pages}{125332} (\bibinfo{year}{2005}).

\bibitem[{\citenamefont{Kyriakidis and Burkard}(2006)}]{unpublished}
\bibinfo{author}{\bibfnamefont{J.}~\bibnamefont{Kyriakidis}} \bibnamefont{and}
  \bibinfo{author}{\bibfnamefont{G.}~\bibnamefont{Burkard}}
  (\bibinfo{year}{2006}), \eprint{cond-mat/0606627}.

\bibitem[{deg()}]{degeneracy}
\bibinfo{note}{There are in fact four degenenerate ground states at $B=0$,
  given by the quantum numbers $S = 1/2$, $S_z = \pm 1/2$, and $L_z = \pm 1$.
  Symmetry prevents mixing between these states and the degeneracy is lifted at
  finite fields. We only consider one of these degenerate states.}

\end{thebibliography}

\end{document}